%% file: main.tex
\documentclass{Interspeech}

\interspeechcameraready

\title{A Novel Deep Learning Framework for Efficient Multichannel Acoustic Feedback Control\thanks{*Work done during internship at Meta Reality Labs Research.}}

\author[affiliation={1,3*}]{Yuan-Kuei}{Wu}
\author[affiliation={2}]{Juan}{Azcarreta}
\author[affiliation={3}]{Kashyap}{Patel}
\author[affiliation={3}]{Buye}{Xu}
\author[affiliation={3}]{Jung-Suk}{Lee}
\author[affiliation={2}]{Sanha}{Lee}
\author[affiliation={3}]{Ashutosh}{Pandey}

\affiliation{Graduate Institute of Communication Engineering}{National Taiwan University}{Taiwan}
\affiliation{Reality Labs Research}{Meta}{UK}
\affiliation{Reality Labs Research}{Meta}{USA}

\email{ywk991112@gmail.com, \{jazcarretao, apandey620\}@meta.com}
\keywords{acoustic feedback, howling suppression, speech enhancement}

\usepackage{comment}

\begin{document}

\maketitle

\input{Sections/0.Abstract}
\input{Sections/1.Introduction}
\input{Sections/2.Methods}

\input{Sections/3.Experimental_Setup}

\input{Sections/4.Results}
\input{Sections/5.Conclusion}

\bibliographystyle{IEEEtran}
\bibliography{mybib}

\end{document}

%% file: Sections/0.Abstract.tex
\begin{abstract}
    
This study presents a deep-learning framework for controlling multichannel acoustic feedback in audio devices. Traditional digital signal processing methods struggle with convergence when dealing with highly correlated noise such as feedback. We introduce a Convolutional Recurrent Network that efficiently combines spatial and temporal processing, significantly enhancing speech enhancement capabilities with lower computational demands. Our approach utilizes three training methods: In-a-Loop Training, Teacher Forcing, and a Hybrid strategy with a Multichannel Wiener Filter, optimizing performance in complex acoustic environments. This scalable framework offers a robust solution for real-world applications, making significant advances in Acoustic Feedback Control technology.
\end{abstract}

%% file: Sections/1.Introduction.tex
\section{Introduction}
Acoustic feedback\cite{waterhouse1965theory, van2010fifty} is a prevalent challenge in audio devices equipped with multiple microphones and loudspeakers, such as hearing aids and public address systems. This phenomenon occurs when the output sound from the loudspeakers is re-captured by the microphones, leading to various disruptive effects such as howling, screaming, and whistling. This issue not only reduces the audio quality but also limits the maximum achievable amplification, thereby affecting the usability of hearing aids, particularly in environments that require high gain settings.

Traditionally, acoustic feedback control (AFC) has been addressed through various digital signal processing (DSP) techniques aimed at identifying and mitigating the feedback path dynamically. One of the most established methods is the use of adaptive filters, such as the Normalized Least Mean Square (NLMS) algorithm\cite{douglas1994family}, which adjusts its coefficients to minimize the error between the predicted and actual feedback. This method, while robust and simple to implement, often suffers from slow convergence rates and can be biased due to the high correlation between the input and feedback signals. Other classical approaches to reduce feedback include delay insertion\cite{van2010fifty, siqueira2000steady, spriet2008feedback, hellgren2001bias, laugesen1999acceptable}, frequency shifting\cite{schroeder1964improvement, strasser2015adaptive}, and phase modulation\cite{guo2012use} to decorrelate signals. Recent advancements like Partitioned Block NLMS\cite{vasundhara2018hardware} and hybrid adaptive filters\cite{nordholm2018stability} enhance convergence and stability.
Moreover, the Recursive Least Squares (RLS) algorithm\cite{van2010fifty} offers improved convergence rates for acoustic feedback control, enhancing overall performance in variable acoustic settings.

In recent years, deep learning techniques have begun to make inroads into the domain of AFC, offering promising results particularly in complex acoustic scenarios where traditional methods struggle. Studies\cite{zhang2023deep, zhang2023hybrid, zhang2024advancing, 10533665} have explored the use of recursive neural network training and hybrid models that combine deep learning with Kalman filters to enhance feedback suppression capabilities. These approaches leverage the ability of deep neural networks to model complex nonlinear relationships and handle variations in the feedback path effectively. However, a significant limitation of these studies is their focus on single-channel settings, which do not align with the multi-channel configurations commonly used in practical audio devices. Moreover, the application of such models is often constrained by their computational demands, making them less suitable for edge devices with limited processing power.

This study proposes a novel deep-learning–based framework to control multichannel acoustic feedback in audio devices, overcoming the limitations of single-channel models and computational inefficiencies prevalent in existing deep learning solutions. We leverage three innovative training methods, In-a-Loop Training, Teacher Forcing, and a Hybrid strategy incorporating a Multichannel Wiener Filter, to optimize our model for complex multichannel environments. Central to our approach is the use of a Convolutional Recurrent Network (CRN)\cite{pandey2024decoupled}, a model that uniquely combines spatial and temporal processing to address multichannel speech enhancement challenges. The CRN model is specifically designed for resource efficiency, characterized by low latency, lightweight architecture, and minimal computational demands, making it ideal for deployment on edge devices such as smart glasses. Here, the minimal delay in processing and the high correlation between feedback and input sources in such devices emphasize the need for robust multichannel processing to effectively mitigate feedback while maintaining audio quality. This model utilizes trainable filters for spatial processing and Long Short-Term Memory (LSTM) networks for temporal dynamics, achieving significant performance improvements over robust baselines with fewer parameters and reduced computational load. Our contributions provide a scalable and effective multi-channel AFC system that is uniquely adapted for real-world applications, offering a substantial advancement in the management of acoustic feedback in audio devices.

%% file: Sections/2.Methods.tex
\section{Methods}

We propose a deep-learning–based framework for controlling multichannel acoustic feedback in audio 
devices with multiple microphones and loudspeakers. Although these devices can take many forms, this 
section focuses on the major elements of acoustic feedback and on three training strategies for our
model.

\subsection{Acoustic Feedback System}
\label{sec:acoustic_feedback_system}

In a device with multiple loudspeakers (indexed by \(j\)) and multiple microphones (indexed by \(i\)), 
acoustic feedback arises when loudspeaker signals re-enter the microphones. Let \(s(t)\) be the 
desired external source (e.g., a user’s speech). Each microphone \(m_i(t)\) contains the desired 
source, and the signals returning from all loudspeakers via their respective feedback paths:
\begin{align}
  m_i(t) = s(t) + \sum_{j} \bigl(h_{ij} \ast y_j(t)\bigr),
\end{align}
where \(h_{ij}\) denotes the impulse response of the feedback path from loudspeaker \(j\) to microphone 
\(i\). The convolution \(\ast\) reflects acoustic propagation 
and enclosure-specific effects. Whenever the loop gain at certain frequencies exceeds unity, howling 
or whistling emerges.

\subsubsection{Default System}

In the simplest default system, each loudspeaker \(j\) outputs the same signal derived from a 
pre-selected reference microphone \(m_{\mathrm{ref}}(t)\). Let \(\Delta t\) be the total system delay, 
\(G\) the amplifier gain, and \(\sigma(\cdot)\) the loudspeaker’s nonlinear response. Then,
\begin{align}
  y_j^{(D)}(t) &= \sigma\Bigl(m_{\mathrm{ref}}(t - \Delta t) \cdot G\Bigr), \quad \forall j.
\end{align}
Because \(m_{\mathrm{ref}}(t)\) itself includes feedback from previous time steps, this setup can cause 
significant howling if not controlled.

\subsubsection{Feedback-Controlled System (MISO Model)}

To suppress feedback, we replace \(m_{\mathrm{ref}}(t)\) with a model-based estimate \(\hat{s}(t)\) of the 
desired source. The model accepts all microphone signals as inputs (MISO: multichannel input, 
single-channel output) and produces one estimated signal:
\begin{align}
  y_j^{(F)}(t) &= \sigma\Bigl(\hat{s}(t - \Delta t) \cdot G\Bigr), \quad \forall j.
  \label{eq:feedback_control}
\end{align}
Since \(\hat{s}(t)\) ideally omits the loudspeaker component from past frames, the feedback loop is 
significantly weakened, preventing re-amplification of loudspeaker signals.

\begin{figure}[t]
  \centering
  \includegraphics[width=\linewidth]{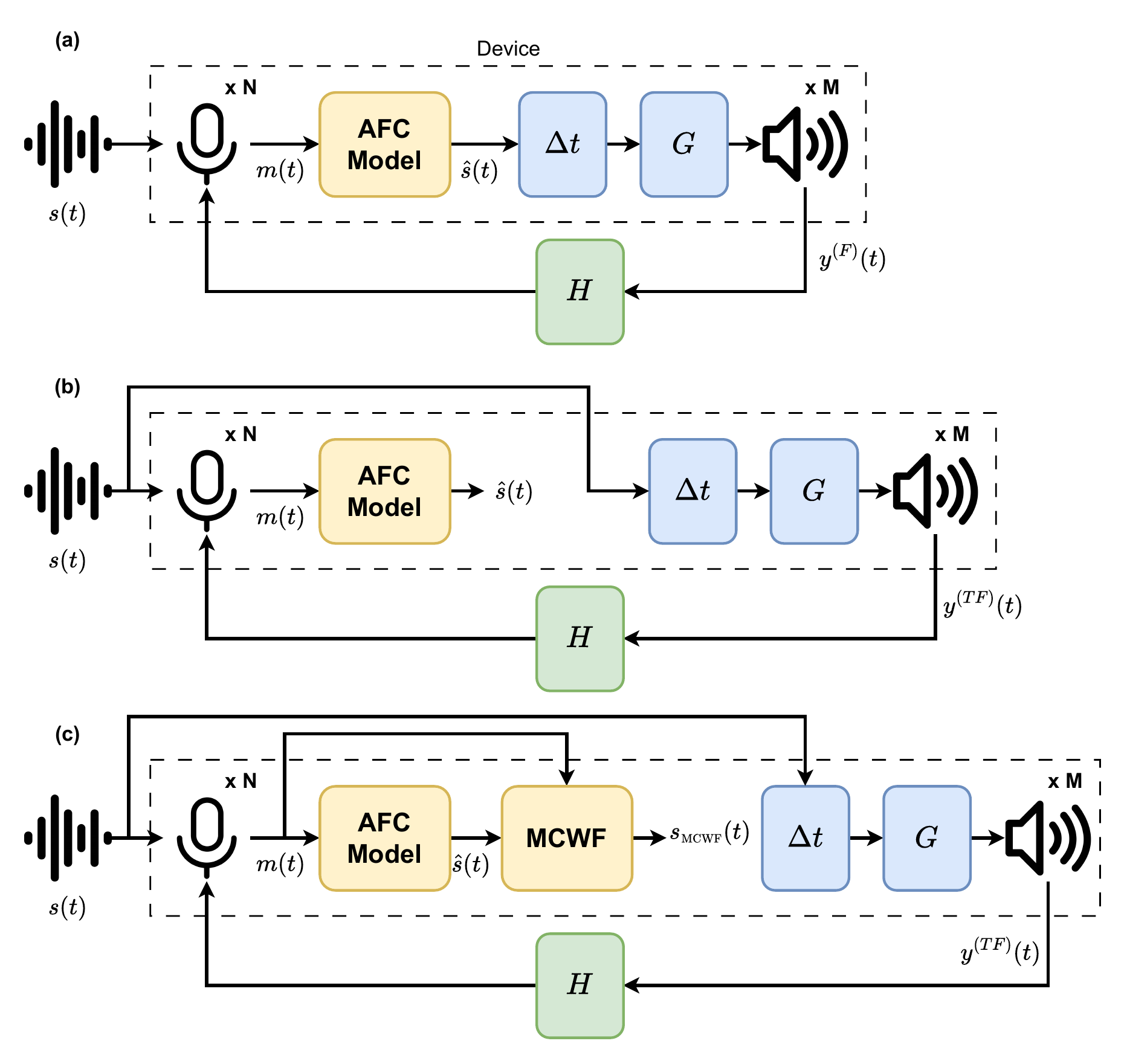}
  \caption{Block diagrams illustrating the three training approaches for acoustic feedback control: 
  (a) In-a-Loop Training, (b) Teacher Forcing Training, and (c) Hybrid with Multichannel Wiener Filter (MCWF).}
  \label{fig:training_methods}
\end{figure}

\subsection{Training Methods}
\label{sec:training_methods}

Below, we describe three strategies for training the proposed MISO model, as depicted in Fig.~\ref{fig:training_methods}. Although each aims to ensure 
that \(\hat{s}(t)\) approximates \(s(t)\) while minimizing feedback artifacts, they differ in how they handle 
the feedback loop during data generation and training. Regardless of the strategy, we define a training 
loss that compares the estimated source \(\hat{s}(t)\) with the true source \(s(t)\). A common choice is the 
Signal-to-Noise Ratio (SNR) loss\cite{le2019sdr}.

\subsubsection{In-a-Loop Training}

This method explicitly simulates the entire feedback process at every time step. The network output 
\(\hat{s}(t)\) is looped back into the system to produce future loudspeaker signals \(y_j^{(F)}(t)\), which 
then influence subsequent microphone inputs.

In this approach, the model generates \(\hat{s}(t)\) at each time step using the current microphone signals. 
These microphone signals include feedback from the loudspeaker outputs at previous time steps. We then 
produce the loudspeaker outputs \(y_j^{(F)}(t)\) by applying the delay, gain, and nonlinearity to \(\hat{s}(t)\). 
These outputs feed back into the microphones for time \(t+1\). After the entire sequence is processed, we 
concatenate all \(\hat{s}(t)\) estimates and compare them with \(s(t)\) using the training loss. This loop simulation closely matches real-time behavior but is 
computationally expensive per sample.

\subsubsection{Teacher Forcing Training}

Teacher forcing assumes that the model completely suppresses any feedback in the loudspeaker signal. 
In other words, it presumes that the loudspeakers play only the pure desired source \(s(t)\) (plus any 
deliberate processing such as delay, gain, and nonlinearity), and ignores the possibility that the model’s 
output \(\hat{s}(t)\) might still contain residual loudspeaker components that could loop back into the 
microphones.

Under this assumption, we take Equation~\eqref{eq:feedback_control} and replace \(\hat{s}(t)\) with 
the ground-truth source \(s(t)\). Thus, instead of relying on the model’s estimate, the loudspeaker output 
is simplified to
\begin{align}
  y^{(TF)}_j(t) &= \sigma\Bigl(s(t - \Delta t) \cdot G\Bigr).
\end{align}
Because \(y_j^{(TF)}(t)\) no longer depends on \(\hat{s}(t)\), the model’s input and output are 
decoupled, and the microphone signals can be generated offline without iteratively 
simulating the model’s feedback. Specifically, we form the microphone input \(m_i(t)\) by
\begin{align}
  m_i^{(TF)}(t) &= s(t) + \sum_{j} \bigl(h_{ij} \ast y_j^{(TF)}(t)\bigr).
\end{align}
During training, the network learns to map these offline-generated inputs \(\{m_i(t)\}\) to the clean 
source \(s(t)\), and the training loss is computed once 
\(\hat{s}(t)\) is produced. This procedure significantly reduces computational overhead, since there is 
no need to simulate real-time feedback from \(\hat{s}(t)\) at each step. However, it creates a potential 
train–test mismatch. In practice, \(\hat{s}(t)\) may contain small errors that re-enter the system 
when the model is deployed, a scenario never observed during teacher-forced training. Consequently, 
while teacher forcing enables efficient offline data generation and can yield strong performance on 
synthetic data, it may reduce robustness once the model faces real feedback conditions.

\subsubsection{Hybrid with Multichannel Wiener Filter}
\label{sec:mcwf_hybrid}

This strategy augments our deep network with a classical Multichannel Wiener Filter (MCWF). The core idea 
is to leverage both the nonlinear modeling ability of deep learning and the spatial filtering capabilities of Wiener-based beamforming in a multichannel setting.

In a multichannel system with \(N\) microphones, let \(\mathbf{m}(t)\) denote the stacked microphone signal:
\begin{align}
    \mathbf{m}(t) 
    = \begin{bmatrix}
        m_1(t) \\
        m_2(t) \\
        \vdots \\
        m_N(t)
    \end{bmatrix}.
\end{align}
In the short-time Fourier transform (STFT) domain, we can write 
\(\mathbf{m}(k, \ell)\) for each frequency bin \(k\) and time frame \(\ell\). The goal of the MCWF is to 
estimate the desired source \(s(k, \ell)\) by forming a linear combination of the microphone channels:
\begin{align}
    s_{\mathrm{MCWF}}(k, \ell) = \mathbf{w}^H(k, \ell) \,\mathbf{m}(k, \ell),
\end{align}
where \(\mathbf{w}(k, \ell) \in \mathbb{C}^{N}\) is the complex-valued Wiener filter for each 
frequency bin and time frame, and \(\cdot^H\) denotes the Hermitian operation.

The optimal Wiener filter \(\mathbf{w}(k, \ell)\) is typically derived by minimizing the mean-squared error 
(MSE) between the filter output \(s_{\mathrm{MCWF}}(k, \ell)\) and the desired source \(s(k, \ell)\). In the 
frequency domain, this involves correlation and cross-correlation matrices. Denoting the 
\textit{autocorrelation matrix} of the microphone signals by
\begin{align}
    \mathbf{\Phi}_{\mathbf{m}\mathbf{m}}(k, \ell) 
    = \mathbb{E}\bigl[ \mathbf{m}(k, \ell)\,\mathbf{m}^H(k, \ell) \bigr],
\end{align}
and the \textit{cross-correlation vector} between the microphone signals and the desired source by
\begin{align}
    \mathbf{\Phi}_{\mathbf{m}s}(k, \ell) 
    = \mathbb{E}\bigl[ \mathbf{m}(k, \ell)\,s^*(k, \ell) \bigr],
\end{align}
the Wiener solution in matrix form is:
\begin{align}
    \mathbf{w}(k, \ell) = \mathbf{\Phi}_{\mathbf{m}\mathbf{m}}(k, \ell)^{-1} \, \mathbf{\Phi}_{\mathbf{m}s}(k, \ell),
\end{align}
assuming \(\mathbf{\Phi}_{\mathbf{m}\mathbf{m}}(k, \ell)\) is invertible. Here \(\cdot^*\) denotes complex 
conjugation.

In our hybrid approach, the deep model first outputs a raw estimate \(\hat{s}(t)\) of the desired 
source, using either in-a-loop or teacher forcing method. Once \(\hat{s}(t)\) is available, we convert it to the STFT domain to obtain \(\hat{s}(k,\ell)\), which serves 
as a reference for the MCWF. Concretely, the MCWF now aims to exploit the spatial information 
from the multichannel microphone signals \(\mathbf{m}(k, \ell)\) and refine \(\hat{s}(k, \ell)\). This 
procedure can be conceptualized as:
\begin{align}
    s_{\mathrm{MCWF}}(k, \ell) 
    &= \mathbf{w}^H(k, \ell) \,\mathbf{m}(k, \ell), \\
    \mathbf{w}(k, \ell)
    &= \mathbf{\Phi}_{\mathbf{m}\mathbf{m}}(k, \ell)^{-1} \, \mathbf{\Phi}_{\mathbf{m}\hat{s}}(k, \ell),
\end{align}
where \(\mathbf{\Phi}_{\mathbf{m}\hat{s}}(k, \ell)\) is the cross-correlation between the microphones and 
\(\hat{s}(k, \ell)\). Here, \(\hat{s}(k, \ell)\) replaces \(s(k, \ell)\) as the desired target in the filter design. 
While this may not perfectly match the ground-truth \(s(k,\ell)\), it provides a guiding signal for the 
Wiener filter that steers the spatial beamformer toward suppressing undesired feedback paths.


%% file: Sections/3.Experimental_Setup.tex
\begin{figure*}[ht]
  \centering
  \includegraphics[width=\linewidth]{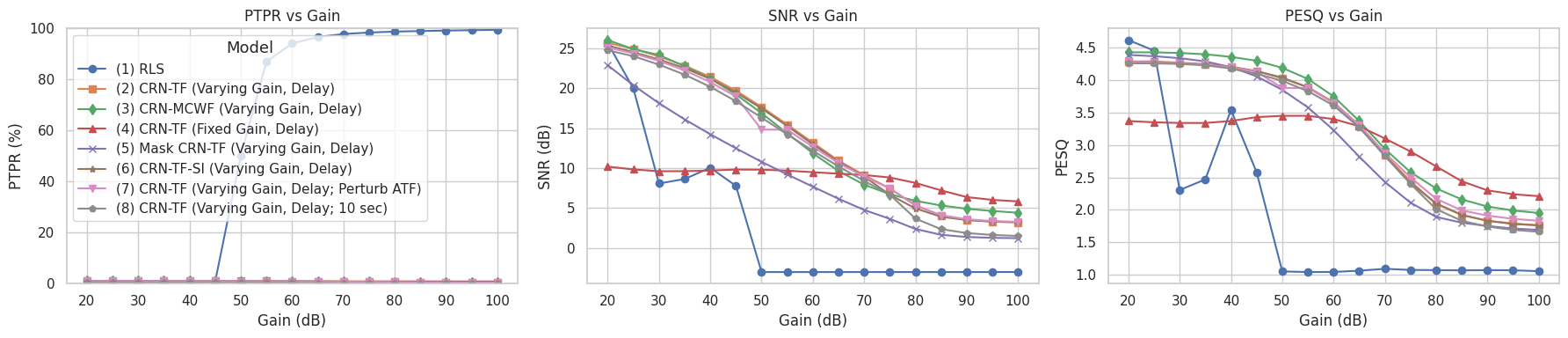}
  \caption{Comprehensive evaluation of acoustic feedback control models showing PTPR, SNR, and PESQ across varying gain levels. Demonstrate the models' capabilities in managing howling, enhancing signal clarity, and maintaining speech quality under different operational conditions.}
  \label{fig:main_result}
\end{figure*}

\section{Experimental Setup}
In this study, we investigate the acoustic feedback problem in multichannel speech data using the Rayban Meta smart glasses microphone array. The input speech is reverberant, sourced from the DNS Challenge clean dataset\cite{reddy2021interspeech}, which provides distinct training, validation, and testing sets. To simulate realistic environments, Room Impulse Responses (RIRs) were generated for 4000 virtual rooms for the training set and 400 rooms for both the validation and testing sets. The microphone array was positioned in 10 different locations within these rooms. The simulation of RIRs utilized the image method with an order of six, considering room sizes ranging from a minimum of 3x3x2 meters to a maximum of 10x10x5 meters. Absorption coefficients varied between 0.1 and 0.4 to mimic diverse acoustic conditions. The source-array distance was varied from 0.5 meters to 2.5 meters. Early reverberation lengths were set at 25, 50, 75, and 100 milliseconds. Each dataset, training, validation, and testing, contains 72000, 3600, and 3600 samples, respectively, with each audio sample having a duration of 2 seconds. The multichannel signals were synthesized using these parameters to study their effects on acoustic feedback in a controlled yet realistic setting.

In the development of our acoustic feedback pipeline, we employ a Convolutional Recurrent Network (CRN) tailored for spatial and temporal processing, based on the frequency domain version of model presented in \cite{pandey2024decoupled}. The CRN model incorporates spatial convolution layers to enhance spatial feature extraction, followed by LSTM layers for temporal dynamics analysis. The process involves channel expansion through spatial convolution, and the resulting features are then processed by LSTM layers for deeper temporal insights. The combined output involves element-wise multiplication of LSTM and spatial convolution outputs to generate the final result. The architecture consists of three CRN layers with channel configurations of 10/20, 20/20, and 20/1, respectively. The model comprises 684K parameters and operates at a processing speed of 82 million MACs per second.

Training settings for the model vary between fixed and variable configurations. Specifically, the amplifier gain is set between 40-75 dB for variable training, and fixed at 75 dB. Similarly, the time delay employed varies between 5-30 ms or is fixed at 8 ms. To emulate the physical constraints of the loudspeaker device, we use a nonlinearity setting of clipping between -1000 to 1000, which effectively mimics the loudspeaker's limitations without inducing howling. During inference, the settings are standardized with a fixed gain and delay. Unless otherwise specified, the gain and delay default to 40 dB and 8 ms, respectively.

As suitable datasets for acoustic feedback transfer functions are not publicly available, we collected our own data using the Rayban Meta smart glasses, an on-the-shelf product. We designed eight specific scenarios to simulate different user interactions with the glasses, including Normal Glassware, Phone Pickup (R/L), Button Press (R/L), Adjust Glasses (R\&L), Cover Ears (R/L), and Grabbing Nose (R). In each scenario, we played back white noise through a loudspeaker and calculated the linear transfer function between the emitted noise and the signal captured by the microphones, capturing the nuances of each interaction scenario. It is critical to note that we assumed a time-invariant system for our experiments, meaning that the feedback transfer function was considered stable and unchanged throughout the time. In this context, 'R' refers to the right hand, and 'L' denotes the left hand.

In the evaluation of our acoustic feedback control system, effective suppression of howling and preservation of speech quality are paramount. We assess the speech quality of the model's output signals using the Signal-to-Noise Ratio (SNR) and Perceptual Evaluation of Speech Quality (PESQ) \cite{recommendation2001perceptual}, with higher values indicating better quality. To detect howling, we employ the Peak-to-Threshold Power Ratio (PTPR) \cite{van2010fifty}, defined as 
\(\text{PTPR}(\hat{\omega}_i, t) [\text{dB}] = 10 \log_{10} \left(\frac{|Y(\hat{\omega}_i, t)|^2}{P_0}\right).\)
Here, \(|Y(\hat{\omega}_i, t)|^2\) represents the power of the candidate howling component, and \(P_0\) is a fixed absolute power threshold set such that \(10 \log_{10} P_0 = 35\) dB. The incidence of howling is quantified by the percentage of frames where \(\text{PTPR} > 0\) dB, with lower values indicating better suppression of howling and improved system stability.

%% file: Sections/4.Results.tex
\begin{figure}[t]
  \centering
  \includegraphics[width=\linewidth]{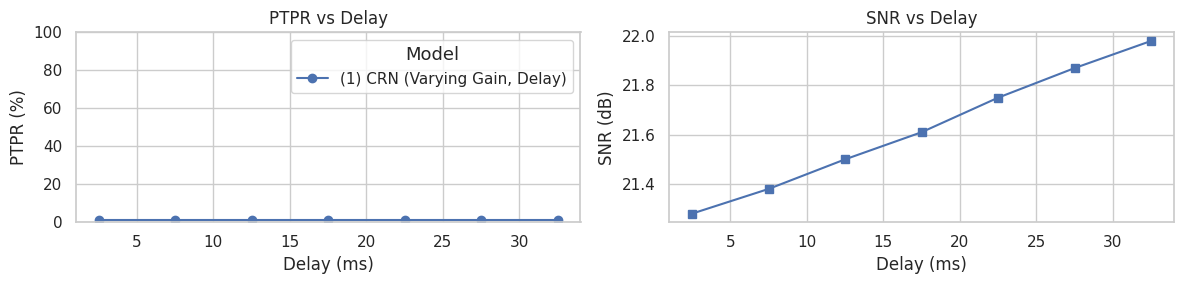}
  \caption{Performance evaluation of the CRN model under varying time delays.}
  \label{fig:delay_performance}
\end{figure}

\section{Results}

In our baseline model, model (1) uses the RLS algorithm, which acts as an adaptive filter and is designed to quickly converge and adapt to rapidly changing system parameters. As depicted in Fig. \ref{fig:main_result}, the RLS algorithm underperforms, particularly in higher gain settings where its propensity to howl is evidenced by poor PTPR scores. In contrast, our proposed model (2), CRN-TF (where TF stands for teacher-forcing), maintains high speech quality with a PESQ score above 4 even at a 50 dB gain. Model (3), which incorporates a MCWF, exhibits superior speech quality, achieving higher PESQ scores. Both models (2) and (3) effectively suppress howling across all gain settings, including gains higher than those in the training set. Although we experimented with In-a-loop training to align training with on-the-fly inference simulation, its unstable training regimen resulted in suboptimal performance across all settings, not shown in the figure.

Subsequent findings demonstrate the performance of various ablation studies on howling suppression and speech quality evaluation using SNR and PESQ scores, as depicted in Fig. \ref{fig:main_result}. When comparing models (2) and (4), where model (2) was trained with gains varying between 40-75 dB and model (4) at a fixed gain of 75 dB, it is evident that the varying gain approach outperforms the fixed gain at lower gains (below 75 dB). However, at higher gains (equal to or above 75 dB), the fixed gain approach yields better performance. This suggests that training with a varying gain enhances model generalization within a broader gain range but is less effective at the maximum gain setting. For models (2) and (5), where model (2) generates the spectrogram and model (5) produces a mask, the spectrogram-based approach shows superior performance in this task. Model (6), implementing system identification (CRN-TF-SI) by using the loudspeaker output as a reference signal, did not outperform the multichannel microphone input method, indicating the challenge of modeling the relationship between feedback signals and loudspeaker output directly. In comparing models (2) and (7), where model (7) included perturbations in the feedback transfer functions by adding Gaussian noise during training. we found that despite introducing noise during training, which created misalignment in the transfer functions between training and testing, model (7) still maintains robust performance in SNR values. This resilience highlights the model's effectiveness in managing SNR despite discrepancies in the acoustic conditions. Finally, comparing models (2) and (8), which tested the effect of extending the inference signal from 2 seconds to 10 seconds, revealed no occurrence of howling and only minor degradation in speech quality. This finding suggests that even with extended inference, the model robustly suppresses howling without significantly compromising speech quality.

Figure \ref{fig:delay_performance} illustrates the performance of our model across various delay settings from 2.5 ms to 32.5 ms. Remarkably, the model effectively suppresses howling at all tested delays, demonstrating robustness beyond the trained delay range. Additionally, we observe an improvement in SNR as the delay increases, suggesting that the model performs more effectively when the feedback signal has less correlation with the input source signal. This trend indicates an enhanced ability of the model to handle increased delays by effectively distinguishing between the source and feedback signals.

%% file: Sections/5.Conclusion.tex
\section{Conclusion}

This study introduced a pioneering deep learning-based framework for controlling multichannel acoustic feedback in audio devices. The framework successfully suppresses howling and maintains high speech quality under various conditions. Notably, our CRN efficiently combines spatial and temporal processing, making it ideal for edge devices with limited capabilities and offering a robust solution for real-world AFC.

%% file: main.bbl
\begin{thebibliography}{10}
\providecommand{\url}[1]{#1}
\csname url@samestyle\endcsname
\providecommand{\newblock}{\relax}
\providecommand{\bibinfo}[2]{#2}
\providecommand{\BIBentrySTDinterwordspacing}{\spaceskip=0pt\relax}
\providecommand{\BIBentryALTinterwordstretchfactor}{4}
\providecommand{\BIBentryALTinterwordspacing}{\spaceskip=\fontdimen2\font plus
\BIBentryALTinterwordstretchfactor\fontdimen3\font minus \fontdimen4\font\relax}
\providecommand{\BIBforeignlanguage}[2]{{%
\expandafter\ifx\csname l@#1\endcsname\relax
\typeout{** WARNING: IEEEtran.bst: No hyphenation pattern has been}%
\typeout{** loaded for the language `#1'. Using the pattern for}%
\typeout{** the default language instead.}%
\else
\language=\csname l@#1\endcsname
\fi
#2}}
\providecommand{\BIBdecl}{\relax}
\BIBdecl

\bibitem{waterhouse1965theory}
R.~V. Waterhouse, ``Theory of howlback in reverberant rooms,'' \emph{The Journal of the Acoustical Society of America}, vol.~37, no.~5, pp. 921--923, 1965.

\bibitem{van2010fifty}
T.~Van~Waterschoot and M.~Moonen, ``Fifty years of acoustic feedback control: State of the art and future challenges,'' \emph{Proceedings of the IEEE}, vol.~99, no.~2, pp. 288--327, 2010.

\bibitem{douglas1994family}
S.~C. Douglas, ``A family of normalized lms algorithms,'' \emph{IEEE signal processing letters}, vol.~1, no.~3, pp. 49--51, 1994.

\bibitem{siqueira2000steady}
M.~G. Siqueira and A.~Alwan, ``Steady-state analysis of continuous adaptation in acoustic feedback reduction systems for hearing-aids,'' \emph{IEEE Transactions on Speech and Audio Processing}, vol.~8, no.~4, pp. 443--453, 2000.

\bibitem{spriet2008feedback}
A.~Spriet, S.~Doclo, M.~Moonen, and J.~Wouters, ``Feedback control in hearing aids,'' \emph{Springer Handbook of Speech Processing}, pp. 979--1000, 2008.

\bibitem{hellgren2001bias}
J.~Hellgren and F.~Urban, ``Bias of feedback cancellation algorithms in hearing aids based on direct closed loop identification,'' \emph{IEEE transactions on speech and audio processing}, vol.~9, no.~8, pp. 906--913, 2001.

\bibitem{laugesen1999acceptable}
S.~Laugesen, K.~Hansen, and J.~Hellgren, ``Acceptable delays in hearing aids and implications for feedback cancellation,'' \emph{The Journal of the Acoustical Society of America}, vol. 105, no. 2\_Supplement, pp. 1211--1212, 1999.

\bibitem{schroeder1964improvement}
M.~R. Schroeder, ``Improvement of acoustic-feedback stability by frequency shifting,'' \emph{The Journal of the Acoustical Society of America}, vol.~36, no.~9, pp. 1718--1724, 1964.

\bibitem{strasser2015adaptive}
F.~Strasser and H.~Puder, ``Adaptive feedback cancellation for realistic hearing aid applications,'' \emph{IEEE/ACM Transactions on Audio, Speech, and Language Processing}, vol.~23, no.~12, pp. 2322--2333, 2015.

\bibitem{guo2012use}
M.~Guo, S.~H. Jensen, J.~Jensen, and S.~L. Grant, ``On the use of a phase modulation method for decorrelation in acoustic feedback cancellation,'' in \emph{2012 Proceedings of the 20th European Signal Processing Conference (EUSIPCO)}.\hskip 1em plus 0.5em minus 0.4em\relax IEEE, 2012, pp. 2000--2004.

\bibitem{vasundhara2018hardware}
Vasundhara, B.~K. Mohanty, G.~Panda, and N.~B. Puhan, ``Hardware design for vlsi implementation of acoustic feedback canceller in hearing aids,'' \emph{Circuits, Systems, and Signal Processing}, vol.~37, no.~4, pp. 1383--1406, 2018.

\bibitem{nordholm2018stability}
S.~Nordholm, H.~Schepker, L.~T. Tran, and S.~Doclo, ``Stability-controlled hybrid adaptive feedback cancellation scheme for hearing aids,'' \emph{The Journal of the Acoustical Society of America}, vol. 143, no.~1, pp. 150--166, 2018.

\bibitem{zhang2023deep}
H.~Zhang, M.~Yu, and D.~Yu, ``Deep ahs: A deep learning approach to acoustic howling suppression,'' in \emph{ICASSP 2023-2023 IEEE International Conference on Acoustics, Speech and Signal Processing (ICASSP)}.\hskip 1em plus 0.5em minus 0.4em\relax IEEE, 2023, pp. 1--5.

\bibitem{zhang2023hybrid}
H.~Zhang, M.~Yu, Y.~Wu, T.~Yu, and D.~Yu, ``Hybrid ahs: A hybrid of kalman filter and deep learning for acoustic howling suppression,'' \emph{INTERSPEECH}, 2023.

\bibitem{zhang2024advancing}
H.~Zhang, Y.~Zhang, M.~Yu, and D.~Yu, ``Advancing acoustic howling suppression through recursive training of neural networks,'' in \emph{ICASSP 2024-2024 IEEE International Conference on Acoustics, Speech and Signal Processing (ICASSP)}.\hskip 1em plus 0.5em minus 0.4em\relax IEEE, 2024, pp. 711--715.

\bibitem{10533665}
------, ``Enhanced acoustic howling suppression via hybrid kalman filter and deep learning models,'' \emph{IEEE/ACM Transactions on Audio, Speech, and Language Processing}, vol.~32, pp. 2828--2840, 2024.

\bibitem{pandey2024decoupled}
A.~Pandey and B.~Xu, ``Decoupled spatial and temporal processing for resource efficient multichannel speech enhancement,'' \emph{arXiv preprint arXiv: 2401.07879}, 2024.

\bibitem{le2019sdr}
J.~Le~Roux, S.~Wisdom, H.~Erdogan, and J.~R. Hershey, ``Sdr--half-baked or well done?'' in \emph{ICASSP 2019-2019 IEEE International Conference on Acoustics, Speech and Signal Processing (ICASSP)}.\hskip 1em plus 0.5em minus 0.4em\relax IEEE, 2019, pp. 626--630.

\bibitem{reddy2021interspeech}
C.~K.~A. Reddy, H.~Dubey, K.~Koishida, A.~Nair, V.~Gopal, R.~Cutler, S.~Braun, H.~Gamper, R.~Aichner, and S.~Srinivasan, ``Interspeech 2021 deep noise suppression challenge,'' \emph{arXiv preprint arXiv: 2101.01902}, 2021.

\bibitem{recommendation2001perceptual}
I.-T. Recommendation, ``Perceptual evaluation of speech quality (pesq): An objective method for end-to-end speech quality assessment of narrow-band telephone networks and speech codecs,'' \emph{Rec. ITU-T P. 862}, 2001.

\end{thebibliography}
